\documentclass[journal]{IEEEtran}

\usepackage{cite}

\usepackage{amsmath}
\usepackage{amssymb}

\usepackage{algorithm}
\usepackage{algorithmic}
\algsetup{linenosize=\footnotesize}

\usepackage{hyperref}

\usepackage{graphicx}
\renewcommand{\figurename}{Fig.}
\usepackage{placeins}
\usepackage{float}

\usepackage[font=footnotesize]{caption}
\usepackage[font=scriptsize]{subcaption}

\usepackage{xcolor}
\usepackage{soul}

\usepackage[acronyms,nonumberlist,nopostdot,nomain,nogroupskip]{glossaries}
\newacronym{iot}{IoT}{Internet of Things}
\newacronym{aal}{AAL}{Ambient Assisted Living}
\newacronym{hci}{HCI}{Human Computer Interaction}
\newacronym{ict}{ICT}{Information and Communication Technology}
\newacronym{qos}{QoS}{Quality of Service}
\newacronym{qol}{QoL}{Quality of Life}
\newacronym[plural=UIs, firstplural=UIs]{ui}{UI}{User Interface}
\newacronym{ECG}{ECG}{Electrocardiogram}
\newacronym{EEG}{EEG}{Electroencephalography}
\newacronym{EMG}{EMG}{Electromyography}
\newacronym{EOG}{EOG}{Electroculography}
\newacronym{AI}{AI}{Artificial Intelligence}
\newacronym{MD}{MD}{medical doctor}
\newacronym{PDA}{PDA}{Personal Assistant Device}
\newacronym{FG}{FG}{focus group}

 \makeatletter
 \def\endthebibliography{%
 	\def\@noitemerr{\@latex@warning{Empty `thebibliography' environment}}%
 	\endlist
 }
\makeatother
\newcommand{\iotAAL}[1]{IoT-based AAL system}

\usepackage[absolute]{textpos}
\usepackage{everyshi}

\begin{document}
	
\title{Internet of Things for Elderly and Fragile People}

\author{Andrea~Zanella$^{1,2,4}$,~\IEEEmembership{Senior Member,~IEEE,}
        Federico~Mason$^{1}$,
        Patrik~Pluchino$^{2,3}$,
        Giulia~Cisotto$^{1,4,5}$,~\IEEEmembership{Member,~IEEE},
        Valeria~Orso$^{2,3}$,
        and~Luciano~Gamberini$^{2,3}$
        
        $^{1}$ Department of Information Engineering, University of Padova, Padova, Italy\\
        $^{2}$ Human Inspired Technology Research Centre, University of Padova\\
        $^{3}$ Department of General Psychology, University of Padova\\
        $^{4}$ National, Inter-University Consortium for Telecommunications (CNIT), Italy\\
		$^{5}$ Integrative Brain Imaging Center, National Center of Neurology and Psychiatry, Tokyo, Japan\\

email: zanella@dei.unipd.it, masonfed@dei.unipd.it, patrik.pluchino@unipd.it, giulia.cisotto.1@unipd.it, valeria.orso@unipd.it, luciano.gamberini@unipd.it
}

\maketitle

\begin{abstract}
This paper discusses the potential of the \gls{iot} paradigm in the context of assisted living for elderly and fragile people, in the light of the peculiar requirements of such users, both from a functional and a technological perspective. We stress some aspects that are often disregarded by the technical community, such as technology acceptability and usability, and we describe the framework and the phases of the current co-design approaches that imply the active involvement of the final users in the system design process. Thereby, we identify a series of design practices to merge technical and fragile people's requirements. The discussion is backed up by the description of DOMHO, a prototypal \iotAAL{} that embodies most of the concepts described in the paper, and that is being deployed and tested in a shelter house for elders, and in an apartment for the co-housing of individuals with disabilities. Finally, we discuss the potential and limits of the current approaches and present some open challenges and future research directions. 
\end{abstract}


\glsresetall

\section{Introduction}
The European Commission recently reported that about one fifth of the European population suffers from some form of cognitive, perceptual or motor disability, making them \emph{fragile} people~\cite{euPersonsWithDisabilities}. 
To address this scenario, the \gls{ict} community is targeting new effective solutions for \gls{aal}, in order to preserve daily living independence of impaired individuals at their future \emph{smart houses}~\cite{Rashidi:2012}.
This has been showed to highly increase the overall \gls{qol} of elderly and fragile people~\cite{Guyatt:1993}~\cite{Beuscart:2014}~\cite{CisottoAESS2018}.

Modern \gls{aal} implementations consist of domestic environments that are equipped with a number of smart, interconnected devices that can continuously collect both health-related data from the users and environmental data from the surrounding space. Such data can then be used locally or sent to remote sites for further processing and storage, to finally become useful information for the persons in need, caregivers and medical doctors~\cite{Dohr:2010}~\cite{Gubbi:2013}. 

In the last years there has been a growing standardization effort that promoted interoperability at all levels among devices from different manufacturers and also with external networks, 
thus expanding the concept of \textit{ambient} to potentially all living environments, both indoor and outdoor. 
A fundamental support for this goal is provided by the \gls{iot} paradigm, thanks to its capacity of connecting different objects in a ubiquitous network. Therefore, care systems can be built by using ready-to-use devices from different providers, rather than closed systems designed for a specific service. The \gls{iot} paradigm is hence a game-changer in this domain, since it can  foster an economy of scale and opening up the market to different players.

Thanks to the diffusion of the \gls{iot} paradigm, an ever larger number of devices and technologies can be used to help elderly and fragile people live independently, despite mental and/or physical impairments.
However, in order to reach their full potential, the sensing devices, the communication protocols and the monitoring strategies (i.e., the entire \gls{iot} framework) need to be chosen in full agreement with the needs, preferences and expectations of final users.
This change has led to a new design viewpoint, generally referred to as \emph{user-centric}, where not only the technical efficiency of the system is taken into account, but also the final users' \gls{qol} is valued, along with the system usability and the user confidence and acceptability of technology.

The user-centric approach requires interdisciplinary skills, ranging from technical to psychological, social and medical skills, in order to account for the multifaceted requirements and needs of these special users and of their complex ecosystem. 
Indeed, these users are likely to encounter difficulties in interacting with a domotic environment, either because of lack of familiarity with digital devices or physical impairments \cite{singh2017ambient}. 
Such impediments may hamper the user confidence in the \gls{iot}-based assisting system, possibly leading to quick disaffection with the technology, or even generating frustration and distress, thereby lowering users' \gls{qol} and  failing the objective of the system design~\cite{Uckelmann:2011}.
On the other hand, a completely customised system cannot be easily replicated to scale, thus resulting in high development costs.

In this work, we aim at providing a more comprehensive vision of the challenges related to the development of \iotAAL{}s for elderly and fragile people.
To this end, we examine the specific needs of individuals with disabilities and elderly people as well as the issues related to the \gls{iot} technology acceptance and usability.
Moreover, we analyze these aspects in DOMHO, an \gls{iot} framework that is under development and test in two real-case scenarios, namely a co-housing apartment for individuals with disability and a shelter house for elders.
Such practical examples help the reader to better focus on the key challenges related to the target users.
Besides, analyzing the best practices related to the current \gls{iot} co-design strategies, we try to propose new ideas that allow to satisfy both the specific requirements of impaired users and the necessity to reduce the capital expenditure for the system realization.
To develop a system that is at the same time affordable and easily replicable, the \gls{iot} solutions need to be designed in order to favor the reusability of the system building blocks.

The remaining of the paper is organized as follows. Sec.~\ref{sec:user} analyzes the \textit{human} component of the system, providing a high-level view of the needs of elderly and fragile users, to identify the aspects of their daily lives where technology can be supportive. Sec.~\ref{sec:technology}, instead, focuses on the \textit{technological} component of the system and presents the most common IoT architecture for e-health services, highlighting some specific requirements that need to be fulfilled to offer a valuable support to elderly and fragile users. In Sec.~\ref{sec:design} we describe the user-centric approach, and list the key features that an \iotAAL{} should exhibit to better address user expectations.
To exemplify the concept, in Sec.~\ref{sec:DOMHO} we describe the approach followed in our prototypal solution, DOMHO.
To conclude, in Sec.~\ref{sec:conclusion} we report some final considerations, and discuss open challenges and research opportunities in the domain of \iotAAL{}s for elderly and fragile people.

\section{Elderly and Fragile People: Needs and Requirements} \label{sec:user}
 \textit{Fragile users} can encompass both younger individuals suffering from physical and/or cognitive disability and older adults affected by cognitive and motor impairments, as they all are likely to share similar needs. 
 
For instance, a community dwelling older resident may be mobility-challenged, as a consequence of a fall, just like a younger adult with motor disability. Therefore, here we adopt a wide perspective by considering both younger and older adults in need as \textit{fragile} users. In this perspective, frailty refers to a dynamic condition in which multiple and interconnected perceptual, cognitive, and physical systems are impaired and weakened, thereby reducing the capability to manage daily activities and/or to cope with everyday stressors. This condition can result into an increased risk of negative health outcomes~\cite{fulop2010aging}.

In the following, we list the most relevant impairments, highlighting how a suitably designed \gls{iot} system can be exploited. A summary is provided in Tab.~\ref{tab:frailty}.
\begin{table}[h!]
\centering
\begin{tabular}{|p{2.4cm}|p{5.4cm}|}
\hline
User frailty &
Technological solutions \\
\hline
\hline
Motor impairments and risk of falling & Deploy domotics and monitoring systems that enable users' independence \\
\hline
Cognitive impairments & Develop gentle reminder systems, promote learnability and memorability of services \\
\hline
Perceptual declines & Implement systems that convert information into different perceptual modalities \\
\hline
Social isolation & Deploy communication services that allow users to interact with relatives and friends \\
\hline
\end{tabular}
\caption{Summary of fragile users' impairments and desirable technological assisting solutions.}
\label{tab:frailty}
\end{table}
\subsection{Motor impairments and the risk of falling}
Elderly people may develop sarcopenia~\cite{barnes2018prevalence}, a well-known disease related to the reduction of the number and size of muscle fibers, which leads to walking difficulties, reduced stability and high risk of falling~\cite{yeung2019sarcopenia}. Similarly, individuals with motor disabilities present a high risk of fall, especially when they have to move from the wheelchair to the bed/sanitary facilities and \textit{vice-versa}. Physical disabilities make these individuals vulnerable, poorly independent and barely autonomous~\cite{owsley2018mobility}, thus needing for continuous assistance (i.e., by professional and informal caregivers). In addition, the reduced muscular strength and frequent arthritis can make it harder to hold and operate devices for a long time \cite{vaportzis2017older}.

Independent mobility in the living environment can be promoted by means of home automation technologies, e.g., automatic doors, remotely controllable windows/shades/lights, portable and movable light switches, smart wheelchairs (capable of autonomous navigation in the environment), and so on.
Traditional monitoring systems feature touch-screen interaction modalities to facilitate older adults \cite{naveh2018age} to control the smart environment. Recently, speech-based interfaces can effectively replace traditional common hand-held devices to relieve fragile users from the effort to hold controls for a long time \cite{masina2019vacs}. 
Technology can also be useful for the prompt detection of falls (e.g., thanks to smart-cameras and smart-watches) and even for preventing them (i.e., using machine learning algorithms). For example, the unobtrusive monitoring of the user conditions and movements, in particular at night or in critical rooms (e.g., bathroom), can raise alerts in case of imminent risk of fall, or trigger alarms in case of dangerous situations (e.g., a person on the floor).

\subsection{Cognitive impairments}
The impairment of one or more cognitive functions is frequent in frail individuals. Processing speed and memory are very likely to be compromised. As a result, fragile users are slower in comprehending, elaborating, learning, retaining and retrieving information. Additionally, attention and executive functions are frequently affected by both the aging process and many neurological conditions. This reflects in the difficulty at keeping the focus of attention for a prolonged time, at inhibiting irrelevant information and at ignoring distracting environmental stimuli. The decline in executive functions also compromises the ability to plan and remember to do things in the near future, e.g., taking medications \cite{naveh2018age}.  

All these conditions are likely to affect not only the planning of the Activities of Daily Living, but also the interaction with control interfaces. However, the cognitive functions needed for the everyday tasks can be supported by the \iotAAL{}, e.g., by guiding the user to unfold such tasks. For instance, the system can automatically set up the temperature of the oven or it can provide step-by-step meal preparation instructions on a large-screen device (\cite{gulla2016adaptive}; \cite{orso2017interactive}). The system may also help the user not to forget appointments or medicine intake, by providing appropriate alerts, which may be either environmental or delivered on a \gls{PDA}.

\subsection{Perceptual declines}

Perceptual abilities are typically affected by age, and they are also frequently impaired in many neurologically diseased people. The decline in visual acuity begins in adulthood and progressively worsens, along with contrast sensitivity \cite{maharani2018visual}.
Likewise, the auditory threshold increases, especially for high frequencies, making it harder to detect and process acute sounds.
Furthermore, it becomes harder to distinguish between targeted speech and background noise, and follow high-paced speeches, thereby compromising the ability to take part in a conversation \cite{czaja2019designing}. 
This mainly reflects on the ability of elderly and fragile users to interact with both people and \glspl{ui} enabling to operate technological services. With this respect, various assistive solutions have been proposed to compensate for perceptual declines.
Typical examples include screen readers applications that convert to sound the text displayed on the screen \cite{olufokunbi2018technology}.

\subsection{Social isolation}
Notably, all the above-mentioned declines increase the risk of social issues, which can in turn negatively impact individuals' affective states (e.g., depression~\cite{lawrence2019hearing}) and \gls{qol}~\cite{tseng2018quality}. Indeed, perceptual decline and reduced mobility may refrain older adults from taking part in social activities, thereby leading to isolation. People with motor and/or cognitive disabilities are likely to experience loneliness, since they mainly interact with their family members and caregivers~\cite{simplican2015defining}. Yet, social relationships are known to have strong influence on the individual's well-being and even health conditions ~\cite{demiris2004}. 
Therefore, besides ensuring the user comfort, the \iotAAL{}s can represent promising tools to maintain and encourage social inclusion. 
In this perspective, \gls{iot} technology can provide a safe platform for facilitating the contacts with relatives and friends\cite{knowles2018wisdom}, e.g., by isolating speech from background noise.

\section{A General Framework for \iotAAL{}s} \label{sec:technology}
In principle, \gls{iot} technologies should make it possible to connect "everything, in everywhere", thus enabling the realization of complex \gls{aal} systems from a variety of elementary building blocks, including sensors, control interfaces, and data management procedures.  

The general framework of an \iotAAL{} is realized through a functional loop which firstly collects signals, commands, and other data from the individuals and the environment they are living in (\textit{observe}). Then, computational facilities are exploited to filter and analyze the available data, integrating, at the same time, information about the context (\textit{contextualize}). In the next phase (\textit{decide}), the processed data are used to correctly classify the individual condition and to identify proper lines of action. Finally, the functional loop provides feedback to the users, either to inform them about their condition or to provide local intervention (\textit{act}). Every phase in the loop is important and requires specific technologies and procedures to be effectively implemented.\\
In the following of this section, we describe the loop phases in more detail, emphasizing the main challenges related to their implementation, with the particular perspective of matching the needs and requirements of elderly and fragile users.

\subsection{Observe}
In this first phase, several sensors, deployed in the rooms of the smart house and worn by the users, allow to pervasively collect heterogeneous data to monitoring the individuals' health and behaviour~\cite{Moosavi2014}. Furthermore, suitable control interfaces are made available to the users, in order for them to easily interact with the system.
To realize a (technologically) effective \emph{observation phase}, it is crucial that the users are compliant with the technology. With this respect, discreteness and sensors' unobtrusiveness are key features to increase compliance and acceptance.

For example, sleep quality is a key sign of the individual \gls{qol} and, thus, it is commonly evaluated to diagnose sleep disorders through polisomnography. The latter is a diagnostic technique that involves recordings by \gls{EEG} (for brain activity), \gls{EMG} (for muscle tone), \gls{EOG} (for eye movements), \gls{ECG} (for heart rate), nasal cannula (for airflow quantification), and pulse oximetry (for the level of oxygen saturation in the blood) over a long-lasting assessment (i.e., one night, after a day of sleep deprivation, most often). However, several prototypes have been recently presented to monitor sleep quality at home, via telemetry stations, such as the one provided by IBM~\cite{Vhaduri2019}. Such an effort is pursued to make future \iotAAL{}s more discrete for the target users, thus promoting their acceptance and compliance.
%
In the same direction, a recent e-health trend sees the active participation of patients to reach a diagnosis: e.g., they are instructed to take pictures of the target body area and to send them via Internet, using their smartphones, to specialized MDs who can formulate a diagnosis~\cite{Rat2018}.
%
It is also a non-trivial question to ensure that sensors are placed, and maintained, in their correct locations: informal caregivers and fragile people should be aware of the impact of incorrect placements of sensors. Even more critically, forgetfulness often characterizing fragile people could make them forgetting to wear sensing and life-saving devices.
%
Finally, privacy on sensitive data has to be ensured in this phase, both in case of individuals taking anatomical pictures for a diagnosis and in case of automatic collection of physiological data. Privacy and security in data transfer to a remote hospital database are very intensively debated open questions~\cite{Abbas2014}.

\subsection{Contextualize}
The data generated in the previous phase are processed, i.e., filtered and cross-correlated, to extract their informative content, which is user- and ambient-specific. 
Relevant multidimensional features could be identified to classify the individuals' condition, based on their current status, past health-related history and environmental conditions.
%
Here, the effective and computationally efficient \emph{aggregation} of different information is a key and critical aspect at this phase~\cite{Ding2019}.
The most recent trend is to feed such big amount of heterogeneous data into deep learning architectures, i.e., complex learning algorithms, to identify abnormal patterns, classify different pathological conditions (e.g., mild or severe) or to predict a prognosis or a medication outcome~\cite{Gomez2016}.
%
These algorithms typically require large computational capacities to be effectively trained and to generalize their classification capabilities over new unseen datasets: therefore, this approach is mostly implemented in the Cloud or at remote powerful servers.
Alternatively, data could undergo only a lightweight local processing~\cite{Samie2016}, to immediately detect relevant events and to reduce the amount of information to be transmitted over the network.

In this phase, the involvement of the final user seems to be more difficult: it might be hard, indeed, to make fragile individuals and their caregivers aware of the processing complexity behind their \iotAAL{}.
However, this is an important question that needs to be addressed to increase user's perception of the system trustworthiness.

Moreover, to increase the accuracy of the algorithms, the sensed data are often integrated with additional data coming from the individuals' electronic health records (EHR): thus, fragile people (or their caregivers) need to provide privacy-related consent to disclose their past health-related data to the \gls{aal} service provider.

\subsection{Decide}
In the third phase of the loop, the information generated during the \emph{Contextualize} phase is exploited to make decisions about what actions and interventions need to be undertaken (in the next phase).
Such decisions can be made by specialized clinicians, \gls{AI}-based expert systems or a combination of them.

Indeed, decision-making is currently recognized as a promising field of application for \gls{AI}~\cite{Shortliffe2018}, thanking to its ability to take into account even tiny correlations between data, as well as their complex and multivariate structure, that a human analyst, even a highly trained \gls{MD}, might not be aware of.
However, the choice of adopting any \gls{AI}-based algorithm to take health-related decisions in \iotAAL{} is still an open question. Indeed, to make reliable decisions, large amounts of data are needed. Moreover, controlling the behaviour of such algorithms requires very specific technical skills.

Therefore, it will be important in the future to have MDs trained to interpret \gls{AI}-based outcomes and to integrate their expertise with the results of those complex algorithms.

\subsection{Act}
Finally, the functional loop is closed by delivering commands to the different actuators in the smart house through the communication technologies, in order to implement the decisions made at the end of the previous phase.

The orchestration of multiple different commands for different actuators have to be properly designed to optimize the communication resources of the network and the energy resources of the battery-powered actuators (e.g., wearables and portable devices).

With this phase, the \iotAAL{} can support fragile people to remember daily drugs intake: recent works proposed the use of a smart TV to remind elderly to take the prescribed medicines on a regular basis. On the other hand, in case any vital is found to be out of its normal range (as outcome of a local computation) an alarm can be generated to alert caregivers or MDs for a prompt intervention \cite{Pires2018}.

The choice of suitable user control interfaces to deliver feedback or messages to the users has to be properly decided, with the active involvement of several kinds of professionals (e.g., psychologists, occupational, physical and speech therapists) in order to optimally set up the interface to effectively communicate with the final fragile or elderly users.

The \emph{act } phase might be particularly important to contrast social isolation of these classes of users. Indeed, tablets, proprioceptive devices (e.g., joysticks) and other interaction devices could be included in the \iotAAL{} to allow individuals to, e.g., take part in serious games, to easily use teleconferencing systems or to take advantage from a remote training session (guided by a human or a virtual therapist)~\cite{Wei2019}. 

Also, suitable interaction devices have to be selected to make users more confident in controlling the system. At the same time, it is desirable that user control interfaces, as well as automated actions or suggested behaviours, are in line with the users' habits: in fact, impairments and ageing cause people to stick more in their habits. Therefore, people compliance to this assisting technology should be promoted by selecting system features as close as possible to their usual behaviours. For example, it is reasonable to place the smart TV in the living room, and smart light switch should replace the conventional ones, i.e., at the same place on the room walls.

\subsection{Technical and functional performance aspects}
To evaluate the performance of an \iotAAL{}, a number of system performance metrics are typically used, including: (i) the delay experienced in transmitting the data collected to the analysis facilities, (ii) the time needed to aggregate such heterogeneous data during the \emph{contextualization phase}, (iii) the computational resources required to accomplish the aggregation (if the aggregation is performed locally), (iv) the user data rate experienced in the transmission of data from the \iotAAL{} to a remote powerful server across the network (if aggregation is performed on Cloud), (v) the relevance of the context-aware information extracted from the aggregated data (locally or in the Cloud), (vi) the efficacy of the action selected to provide the user with a proper feedback, (vii) the overall loop execution time.
Although other performance indexes could be considered, they could be generally categorized in three main classes of metrics: time-related, computation-related, and \gls{AI}-related.

For instance, the overall loop execution time can have a critical impact for some applications, which may require latencies in the order of (less than) hundreds to tens of milliseconds ~\cite{Cisotto2019}.
Then, decision-making is not a trivial problem, especially when the human user is included in the system functional loop, providing much larger variability in the dataset to analyze. Here, the \emph{contextualize phase} and the \emph{decide phase} have to process such a heterogeneous big dataset to provide the proper feedback at the right time: its effectiveness for the users highly depend on these aspects. 

However, all the above-mentioned technology-related metrics do not ensure users' satisfaction: even the most efficient system proves to be under-performing when it does not take into account users' preferences and individual needs.
Therefore, additional user-related metrics have to be taken into account to evaluate the performance of an \iotAAL{}.

As an example, the system should show easy and immediate control by all target users, thus improving the level of \emph{accessibility} and, consequently, promoting technology acceptance. Importantly, accessibility level is boosted by the capability of the system to rapidly recover from users' errors, thereby preventing frustration, especially in case of elderly and fragile users. In addition, it can also prolong their willingness to use the system over time.

Moreover, a proper limited selection among all available services might be activated for each individual user: this may further increase the system usability, by decreasing the complexity of the control interface.

Then, if not properly designed, technology could easily yield individual isolation, as well as push fragile users to become fully system-dependent: therefore, another important user-related metric is defined by the capability of the system to represent a supportive tool for them, without reducing their independence or discouraging them to engage in social relationships. 

\section{A User-Centric IoT Design Framework}\label{sec:design}

The literature has recently reported attempts to match users' needs, values, habits and, generally, lifestyles, when designing new \iotAAL{}s~\cite{Domingo:2012, Miranda:2015, Soro:2017}.
%
Interestingly, Pires \emph{et al.}~\cite{Pires2018} implemented an \gls{iot} platform to take care of elderly at their home, where a smart TV was chosen as a means to convey additional health-related information to the individual, in an unobtrusive way (e.g., as additional commercials).
Both in~\cite{Mangano:2015} and~\cite{Hussain:2015}, authors aimed at building a user-centric e-health platform to assist elderly and people with disability in \textit{smart cities}.
In~\cite{Lopes:2014}, the \gls{iot} architecture, itself, has been modified, with an additional layer, to address the specific needs of people affected by diverse disabilities.
\cite{Miranda:2015, Soro:2017} claimed the potential of their systems to progressively learn users' behaviors, adapting services accordingly.
Finally, some standardization effort to provide effective user-centric \gls{iot}-based \gls{aal} solutions has been promoted in~\cite{Domingo:2012} as a major challenge for the actual spread of \iotAAL{}s in healthcare.
Indeed, system flexibility and learnability have come up as key features of the user-centric \gls{iot} design, which could significantly increase the level of system usability, besides efficiency and security alone ~\cite{Thomas:2016}.

In what follows, we start from the most recent findings of the scientific community to draft a set of best practices for \iotAAL{}s design.
Such practices aim at combining the human and the technical requirements, to move from a techno-centric perspective, where {technical performance} have a key role, towards a \textit{user-centric} approach, where \textit{multi-modality} and \textit{customization} allow the system to adapt to the user.
Indeed, the ambitious objective of the new generation of \iotAAL{}s is to support high user \gls{qol} while maintaining technical efficiency and flexibility to generalize the same solution to different cases. We can anticipate that \textit{modularity} plays a critical role to break down the overall system into elementary modules that can be customized and reused in the future, at the same time.

\subsection{Co-Design}~\label{sec:co-design}
\gls{iot} technologies can be an effective support for elderly and fragile users to accomplish the Activities of Daily Living with a higher degree of autonomy. However, if the services are not appropriately designed, technology may become an additional barrier to people's well-being~\cite{de2016accessibility}.

Co-design consists in actively involving the final users in the development of the target system. Complying with technical guidelines may not be enough to realize a system that can fully match people's needs. Indeed, developers and designers do not share the standpoint, the expectations and the preferences of end users. Previous findings showed that even caregivers can fail to identify the tastes of their assisted older adults~\cite{spagnolli2012involving}, thereby highlighting the importance of user active participation in the system development. Hence, user needs and preferences are kept at the center of the entire design process, and special care should be taken on the methods deployed to engage them. 

The current co-design approaches develop along the phases depicted in Tab.~\ref{tab:Co-design}. The first steps consist in defining goals, scenarios, and the identification of users’ needs and requirements, e.g., through \gls{FG}s. Based on this information, the developers can create mock-ups of the main system functionalities (e.g., using fake cameras and wearables to test users' acceptability of such technologies), which will be used to choose the most suitable \glspl{ui} and sensing/actuation devices to be integrated in the final system. At this point, the different components can be assembled into the customized \iotAAL{}.
The users are then trained on what are the system services and how to access/control them. This training should be as natural and automatized as possible, and should gently and progressively drive the users to acquire familiarity with the \glspl{ui} and the system services by means of first-use instruction paths, examples, suggestions etc., as today customary for many apps and video games. After the actual usage, users are invited to provide a feedback regarding their level of satisfaction and overall user experience. Such a feedback can then be used to fine-tune the system, while enriching the knowledge about the acceptability of certain technical solutions.  
 

\begin{table}[h!]
\centering
\begin{tabular}{|p{2.4cm}|p{5.4cm}|}
\hline
Methods &
Outputs \\
\hline
\hline
Defining goals and scenario &
Set of information on which the \gls{FG} or other methods (e.g., Affinity Diagram) of ideas and knowledge elicitation will be based \\
\hline
\gls{FG} and Elicitation of ideas and identification of users' needs and requirements\\
\hline
Mock-up                                 & Development of mock-ups of UIs and system prototypes for user testing \\
\hline
Testing & Evaluation of system performance, usability, accessibility, acceptance and user overall experience \\
\hline
System/UIs iterative co-development & possible modifications of systems/prototypes considering their performance and users' feedback followed by further testing sessions \\
\hline
\end{tabular}
\caption{The phases of a typical co-design process. To note, methods can be always adapted to meet the skills of fragile users.} 
\label{tab:Co-design}
\end{table}

Traditional co-design methods, e.g., \gls{FG}s, have to be adjusted to meet the skills of the users. For instance, in a \gls{FG}, the difficulty of older adults to abstract concepts has to be taken into account and pictorials can be included to facilitate the discussion. For the same purpose, video clips can be used to overcome linguistic difficulties, i.e., technical jargon, and to help users conjure up the topic of interest~\cite{lindsay2012engaging}.  
The same considerations are applied also in the evaluation phase, where researchers adjust testing situations and task requests while keeping in mind the low level of familiarity of the specific users to technology and their fragility \cite{orso2015involving}. The system developers should also arrange the experimental setting to resemble a typical, thus familiar, living environment and to frame the experimental situation into a sequence of more natural and easier tasks \cite{dickinson2007approaches}.  

\subsection{Multi-Modality}
The outcomes of the co-design phase are the basis for all the rest of the development choices, including the introduction of multiple interaction modalities. 
Multi-modality consists in providing multiple perceptual channels (e.g., touch, vision, hearing) by which the user can interact with the same smart device. 
In other words, multi-modality provides such a redundancy of controllability that ensures people to be able to interact with the \gls{iot} system in the way that better suits their characteristics~\cite{obrenovic2007universal}.
This is particular relevant when considering people with cognitive and/or motor disabilities.
Besides overcoming perceptual impairments, distributing information in multiple formats allows the system to adjust to the changes of the user's health condition, thus prolonging its utility and usability~\cite{de2016accessibility}.

For instance, touchscreen interfaces of smartphones and tablets are commonly used to control domotic devices in smart houses~\cite{kuhnel2011m}.
However, fragile people with muscular weakness, tremors or impairments in eye-hand coordination and fine gestures can hardly control the system through tapping, sliding, multi-touch gestures and timed taps.
Alternatively, voice-based interfaces can be offered, e.g., to control opening/closing of doors and powering on/off devices.
Nevertheless, they typically rely on strict command-pause-command timings that could make fragile users struggle with, because of their difficulties in correctly, promptly and loudly pronouncing predefined commands (most likely, users tend to utter generic sentences using a more natural language).

Providing a system that integrates both touch and voice-based interfaces (as well as other perceptual channels) could be the only way to ensure the full usability of the implemented solutions.

\subsection{Customization} 
With customization, we mean the possibility for a user to \textit{autonomously} change the system settings according to her/his preferences.
A recent review by Dood \textit{et al.}~\cite{dodd2017designing} has underlined several aspects that have to be considered, including the interface design (e.g., text and icon features), the input controls (e.g., voice, eye-gaze), the usage of natural language (i.e., avoiding technical jargon). 

The interfaces for controlling the system should be devised to be flexible enough to match the level of ability of every single user.
Besides personal preferences of using a specific interaction modality, the user should be able to control the sensitivity of the input and the intensity of the output signals.
For example, the end-user should be free to set the desired loudness of the audio feedback or the intensity of the tactile feedback.
Similarly, the contrast ratio between the text and the background, the size of the textual labels should be in line with the perceptual (e.g., visual ability) and cognitive characteristics of the end users \cite{flores2019geo} and adjustable whenever necessary, to maximize information readability.
Furthermore, the \glspl{ui} has to enable the user to customize the amount and the organization of the information visualized on the display to prevent cognitive overload, which may lead to confusion and frustration \cite{sharma2020three, pirzada2018sensors}. Thus, a fragile user may decide to reduce the information on the display to a few essential pieces (e.g., only icons).

Clearly, such a customization feature should be offered in an intuitive and easy-to-access format, without necessarily requiring to go through complex dashboards with long lists of adjustable parameters.
Instead, the system should be able to self-tune some parameters, based on the user's setting of others.
For example, if a user opt for vocal rather than touch \glspl{ui}, the system may automatically set the preference from a loud audio feedback to acknowledge the reception of other commands, when required. 

\subsection{Data privacy and security}

Since assistive \gls{iot} systems deal with users personal data, privacy and security are of special relevance. The security of an \gls{iot} framework is based on \textit{confidentiality}, \textit{authentication} and \textit{authorization}~\cite{Lin:2016}.
The full adherence to these concepts may be very challenging: indeed, the \gls{iot} paradigm intrinsically makes it possible to access the system's information from multiple points.
Moreover, \gls{iot} technologies present severe computational and energy constraints; developers may not be able to develop a security procedure that requires a large amount of such resources~\cite{yang2017}.

Finally, \gls{iot} systems typically include many heterogeneous, often off-the-shelf, devices with different features and their own commissioning and security procedures.
The latter can be very challenging for the users to master, it can be difficult to configure an unique security protocol over all of the system components~\cite{zhang2014} and the future expansion of the system with new elements, or the replacement of obsolete ones, might be a non-trivial issue.

It is essential that an \gls{iot} system for elderly and fragile people implements a full-fledged security architecture that guarantees the protection of the user data.
At the same time, the security procedure should be simple and intuitive, avoiding complicated commissioning mechanisms and the need for the user to enter and memorize complex passwords. In this respect, biometric systems based on visual, acoustic, or kinetic signatures appear as simple and effortless authentication solutions, and thus particularly suitable for fragile individuals.
To provide a centralized control of the different devices' security mechanisms, the system should support a \textit{security manager} entity that integrates and wraps different protocols, in order to offer a single authentication method to the user. This aims at both reducing the cost of the \gls{iot} architecture and facilitating users to attend security procedures.

We highlight that, compared to healthy individuals, fragile people are more willing to allow smart devices monitor their status and collect their sensitive data for third parties management and analysis, if this helps them to maintain independent living for a longer time. However, this attitude may be detrimental, since the users may disregard the most fundamental security procedures, failing to see the potential impact of security breaches for their privacy and safety.

\subsection{Technology trustworthiness}
To maximize the usability of all system functions, a user-centric framework has to ensure that users fully trust the implemented devices.
In this perspective, it is important not only to involve fragile users in the operational stages of the \gls{iot} platform, but also to clearly present the goal of the application and the related security properties in a comprehensible way~\cite{hochleitner2012making}.
For instance, a system that is designed to avoid unintended actions and capable to support the user in the quick recovery from errors is considered more reliable, thereby trustworthy~\cite{kainda2010security}.
Another key factor is the accessibility, i.e., the capacity of the system to enable a natural interaction, without requiring the user to think about detailed security aspects~\cite{hochleitner2012making}. 
Moreover, developers should give great importance to the system capacity to adapt to different routines and lifestyles, which can increase the user confidence.

Even paying great attention to these aspects, the introduction of new technologies into people's life may be critical.
It should be considered that interaction devices are strongly context-specific, especially for the elderly.
Despite the availability of technical solutions, people are likely to stick to their habits (e.g., users will not use the intercom if they are not in the living room~\cite{kim2015understanding}).
Moreover, the continuous data gathering of \gls{iot} sensors may generate a sense of discomfort in people.
Hence, even when system security is always guaranteed, a user may distrust smart devices and avoid to use them~\cite{martin2008}.
Finally, certain individuals prefer to be assisted by devices with a limited amount of computational intelligence~\cite{takayama2012}.
Recent advanced smart home applications explore techniques based on machine learning to identify users' behaviors and to take actions accordingly.
This may make people think they lack control over technology and, consequently, they might stop using it~\cite{norman2014}.

In this perspective, it is fundamental to make the users aware of all processes by which their data are gathered.
At the same time, the automation level of the overall system should be adjustable.
The system developers should pay attention to both these aspects, to ensure that the target users fully trust the implemented solutions.
The perceived usefulness and reliability of \gls{iot} technologies are predictors of the willingness to use the system, thereby being associated to its success or failure~\cite{park2017comprehensive}.
Thus, user confidence of fragile users (and their informal caregivers) increases if they are fully informed about the specific benefits and features of the \gls{iot} system.

\subsection{Modularity}
Although the characteristics of elderly and fragile people call for highly specific implementations, it might be not feasible to conduct an \textit{ad hoc} co-design procedure for every individual user.
Indeed, in a realistic scenario, the different system components are developed separately and integrated later, into a unique architecture.
This may lead to a very heterogeneous \gls{aal} environment, which can be hardly adapted to new different contexts.
Hence, the only way to promote both specificity and flexibility is by dividing the system into multiple compatible modules. 

System modularity enables new applications to be run via continuous aggregation and integration of individual and elementary units on the top of a basic structure.
Each unit implements a single function of those required by the system, e.g., human interfacing, environment sensing, data computation, or decision actuation. 
Hence, the number and the kind of system components change depending on the scenario and the considered class of users.

Control nodes may be added to the framework to better synchronize and manage the activities of the different modules.
Moreover, they could facilitate the inter-operability between different parts of the system, as well as different security and communication protocols.

Thanks to modularity and to control nodes, the user does not need to be aware of the specific communication architecture that interconnects different system components. For instance, human-machine interactions can be operated by a limited number of control interfaces, i.e., those mostly suited to the user's preferences. The latter are directly connected to the control units, hence allowing the user to govern multiple devices and functions, irrespective of their specific procedures and operations. 
Thus, elderly and fragile people need only to familiarize with a single general-purpose interface that can allow them to control and customize the entire system. User confidence and satisfaction are expected to be enhanced by modularity, as user's needs are constantly supported by proper modules, easily integrated in the basic platform.

Finally, modularity may allow the \gls{iot} system to achieve the market level. indeed, integration of customized elementary modules and the scalability of the prototype solution can make the system effective for new classes of users.

\section{The DOMHO System} \label{sec:DOMHO}
DOMHO is a prototype \iotAAL{} that has been originally designed by the authors and is being under development within the framework of a 3-year project\footnote{Project No. 10066183 titled “Sistema domotico IoT integrato ad elevata sicurezza informatica per smart building”, funded under the ``POR FESR 2014-2020'' Work Program of the Veneto Region (Action 1.1.4).}.
The aim of DOMHO is to provide an integrated domotic system for smart buildings to enhance the \gls{qol} of elderly and fragile people, and their caregivers~\cite{gamberini2018cyber}.

DOMHO exploits \gls{iot} hardware solutions to enhance well-being, safety, autonomy, and independence of the target users. The project consortium comprises three Universities, lighting and sensors companies, software and hardware developers, and automation industries. The project operates in two real-world scenarios: a co-housing residential apartment and a shelter house. In the first scenario the final users are individuals with disabilities (i.e., mainly severe motor impairments and in some cases comorbidities with cognitive disabilities). In the second one the end users are elderly people (i.e., older than 65). Finally, in both scenarios caregivers are considered as relevant stakeholders.

\subsection{Co-design}
The development of the DOMHO system is being followed the guidelines provided in Sec.~\ref{sec:design}.
In the first two years, the project implemented a participatory design approach to identify, adapt, and develop hardware and software solutions based on user needs, expectations, and desires. This information was gathered by using well-established methods (e.g., \gls{FG}~\cite{parker2006focus,singh2017ambient}) for eliciting and organising ideas and knowledge. 
The \gls{FG} sessions were conducted by HCI experts, involving end-users both in the co-housing and shelter house scenarios. In two \gls{FG} sessions, the users from the co-housing scenarios participated.
Five users (3 females,  mean age = 39,6) and five stakeholders (i.e., 3 professional caregivers,  2 family members) took part in the first \gls{FG}. Four end-users (3 females, mean age of 41 years) and four stakeholders participated in the second \gls{FG} (i.e., 2 professional caregivers,  2 family members). In regard of the shelter house, a \gls{FG} involved two users (1 female, mean age of 75 years) and nine stakeholders (i.e., nurses, social and health professionals). All \gls{FG}s were audio-recorded for allowing the analysis of the transcriptions. 
According to the co-design practices described in Sec.~\ref{sec:co-design}, the technological devices that best served the needs of the two user groups were selected.
This approach enabled to highlight similarities and eventual differences in the needs and expectations between the two end-users' categories.

For all system applications, the field trials (in both the scenarios) have been preceded by laboratory tests concerning the technological prototypes that will constitute the fully-integrated \gls{iot} final system.
%
Particularly, an iterative process is being followed: HCI experts from academia have tested the preliminary prototypes with end users, in a laboratory setting. Specifically adapted tools are employed \cite{orso2015involving}, including questionnaires, computer-supported video-analysis \cite{silva2018ambient} and performance metrics (i.e., time on task, task-accuracy, number of interactions), to collect information about user experience, system accessibility and usability. Designers and developers are now able to identify and fix possible issues in the system, accordingly.

\subsection{Preliminary assessments}
Two experimental sessions were carried out to test two prototypes of \gls{ui}, i.e., a smartphone and a tablet, for controlling the home automation.
Eight individuals with disability (i.e., 4 in each session) and four caregivers (i.e., 2 in each session) were involved. 
In each session, in the first screen of both interfaces, intuitive text/labels of a selection of environments (e.g., kitchen, living room, etc.) and scenarios (e.g., night) are presented. Users had to select one environment by pressing the corresponding icon. In the second screen, they could control specific smart objects of that room by selecting the corresponding image/label (e.g., lights, roller shutter, TV).
Individuals with disabilities were asked to perform a series of tasks (e.g., turn on and off smart lights or changing their colours) utilizing both interfaces. Caregivers, instead, used only the smartphone interface for carrying out a different task, i.e., controlling several smart devices simultaneously.
At the end of each session, participants had to fill questionnaire about the perceived utility, accessibility, and aesthetic features (e.g., information organization) of the prototypes.

Along with the overall positive evaluation of the interfaces (gathered from the questionnaires), the analysis of the errors made by the users during the tests highlighted some design issues. 
For instance, in future prototypes it will be necessary to increase the dimension of some icons to better deal with the reduced motor and visual abilities of the target users (e.g., home button).

Besides, an additional series of tests was performed to assess the suitability of a Voice Actuated Control System (VACS) to control smart devices (i.e., lights, plugs, TV) in a living laboratory.
Findings proved the feasibility of such means of interaction for individuals with severe motor disabilities and mild cognitive impairments\cite{masina2019vacs}.
In fact, despite the consistently high number of trials necessary to successfully implement some of the commands through the VACS unit, all users were able to perform the administered tasks. Thus, voice control has been included among the \gls{ui} modalities for DOMHO.

\subsection{Multi-modality, customization, and technology trustworthiness} 

The two abovementioned \glspl{ui}s, i.e., the smartphone and the tablet, have been developed using slightly different interfaces for users, caregivers and installers/maintainers.
The devices have been equipped with different operation modes to meet the needs of various categories of individuals, following the principle of multi-modality described in Sec.~\ref{sec:design}.
Moreover, specific operation modes can also allow to verify permissions and to adjust parameters by the installers/maintainers when needed.

An example of multi-modal interface implemented in the DOMHO system consists in the lighting control, which will be possible either through Bluetooth wall switches or virtual buttons on the touchscreens of smartphones and tablets.

Technology trustworthiness is one of the aspects that DOHMO will consider in the future experimental trials: whenever a new system component has to be designed or selected, accessibility and usability principles (i.e., multi-modality control, natural interaction) will be taken into account to ensure the successful introduction of such new technology and its adoption over time.
Nonetheless, a recent paper \cite{knowles2018wisdom} highlighted three main factors that may hamper the ability of older adults to use technologies: \textit{responsibility}, elderly are not comfortable in performing tasks that were handled by professionals in the past; \textit{values}, older individuals may not use technologies that are perceived as a replacement of something valuable to them; \textit{cultural expectations}, aged people perceived themselves as a population segment for which not using technologies seems culturally acceptable. Thus, this specific contextual environment and expectations have to be carefully understood and addressed before introducing a novel technology to support fragile people.

\subsection{Safety and privacy protection}
\begin{figure}[!htpb]
\centering
    \includegraphics[width=0.5\textwidth, trim={0cm 0.5cm 0cm 0.5cm}, clip]{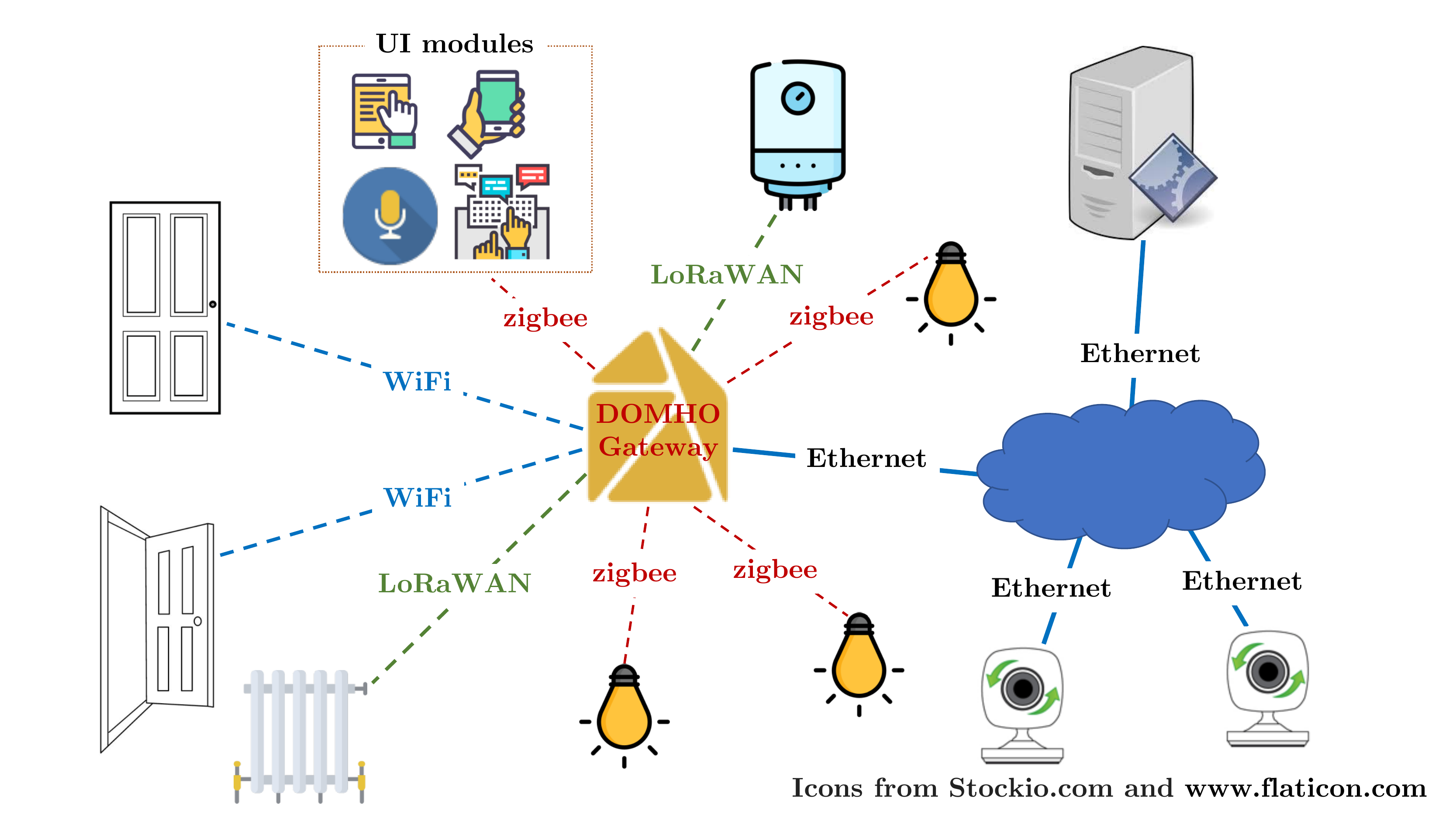}
\caption{The DOMHO system.}\label{fig:domh}
\end{figure}

The target architecture of the DOMHO system is sketched in Fig.~\ref{fig:domh}. In its final implementation, the DOMHO system will ensure safety by 3D smart cameras installed at specific locations (e.g., bathrooms, bedrooms, corridors). These devices allow the early detection of individuals' falls, enabling a prompt assistance. To preserve the user's privacy and intimacy, however, the images collected by the videocameras will not be shared with other devices, rather they will be processed locally to extract only those useful information to protect the individual safety. Locations for the installation have been already identified during the \gls{FG}s when the participants unanimously reported the places in which they are more afraid to fall or they have actually fallen. Participants generally stated that they would feel more secure having the support of the smart camera.
Safety specifications have been defined according to the scientific literature and the discussions during the \gls{FG} sessions.
In particular, machine learning algorithms are being developed in order to predict and eventually prevent falling of elderly persons: they typically analyze the position of the body while individuals move throughout the environments or stand from the bed, and also the body posture during sleep.
Moreover, the installed cameras will communicate with the lighting and automated access systems, allowing the user to safely move during the night-time (for instance, to get to the toilet). Furthermore, both the cameras and the automated doors will exchange information regarding the physical location of each user to prevent accidents due to the phases of door opening/closing. The installation of all safety devices is consistent with the users' daily routines, so as to promote user trust in the overall system.

\subsection{Modularity}
In the next phases of the project, all devices we previously described will be mutually connected in a centralized system. In particular, a smart gateway will provide communications among all system components, managing the data exchange between the sensors and the Cloud, and bridging different communication technologies. The smart gateway, indeed, will support software wrappers to integrate different off-the-shelf products and translate the standardized commands received by the centralized orchestration and management module into the specific formats required by the end devices. The central controller will run on a server and, among other services, it will support the security framework that will effectively work on all system components. Hence, modularity will be promoted so that the system will efficiently suit the characteristics of each user.

\section{Conclusions} \label{sec:conclusion}

The design of \gls{iot}-based \gls{aal} systems to assist elderly and fragile individuals is particularly challenging.

We claim that the potential of \gls{iot} for elderly and fragile people can be fully expressed only by adopting a \textit{fully modular design principle}, which extends from the identification of users' needs and service requirements (co-design with all stakeholders and target users, and system customization) to the integration of different individual system components, i.e., modules. Indeed, modularity should be realized at \textit{all levels} of the system, in order to allow for full customization, in line with co-design outputs, while guaranteeing re-usability of the components and, hence, economy of scale.

These principles have been followed in the design of the DOMHO system, as introduced in Sec.~\ref{sec:DOMHO}. Compared to existing pilot studies, DOMHO merges the co-design approach together with a modular system architecture, which entails multiple network technologies (Ethernet, ZigBee, LoRaWAN), and a variety of user control interfaces. The different components are seamlessly interconnected by a central smart gateway, which makes it possible to easily change sensors, communication protocols, and control interfaces to better adapt to the use-case characteristics and the users' requirements. This will make it possible to reuse most of the software and hardware components of this system for new scenarios, paving the way of using DOMHO for other classes of elderly and fragile people needing assistance.

\subsection{Open research challenges}

Even though DOMHO represents a step forward compared to exiting solutions, much has still to be done to realize the vision of a fully modular user-centric \gls{iot} system architecture, suitable for elderly and fragile users needs.  In what follows, we list a number of open research questions and design challenges that have to be addressed to reach such an ambitious goal. 

\paragraph*{Interoperability}
Supporting efficient and reliable health-care services from hospital to home, to rural areas, and even in mobility, is still an open challenge \cite{Cisotto2019}.
The system needs to support and exploit heterogeneous communication protocols, sensors, and software modules to cope with the variability of environments and service requirements, but most of off-the-shelves products are scarcely inter-operable, thus forcing developers to invest time and money in customized solutions, with limited re-usability.

\paragraph*{Standardization}
Despite the massive effort of different standardization bodies towards the definition of guidelines that can ease interoperability among \gls{iot} technologies, still general consensus has not been reached yet~\cite{Badawi2018, Eom2017}.
In particular, standardization should address the format to be used for data exchange and  storage, which is fundamental to enable knowledge sharing among care institutes, hospitals and research institutes~\cite{Islam2015, Rodrigues2018}.

\paragraph*{Security}
Cognitive impairments may affect elderly and fragile people, thus making it difficult to manage different security procedures for all devices in the system.
Therefore, an open challenge is to devise security frameworks that can support single-sign-on access to the system, possibly using biometrics or other physiological signals~\cite{Natarajan2018}.

\paragraph*{Scalable co-design}
The modularity principle should include the co-design phase, which may be performed by means of dedicated devices to collect feedback from the users.
These operations are currently demanding, requiring HCI experts to run \gls{FG} sessions, which require transcriptions to be, following, analyzed. 
The development of more agile tools, e.g., adequately adjusted surveys, of the user's satisfaction could speed up the co-design and testing phases, thus greatly reducing the design time and cost.

\paragraph*{\gls{AI}}
The potential of \gls{AI}, i.e., machine and deep learning algorithms, in \iotAAL promises to be huge, but it is still object of several ongoing studies. 
Such methods may automatically adapt the configuration of the system interfaces to the user preferences, and process the data collected by sensors to provide specific services, as pursued with the DOMHO project.
Although they have already showed promising results, many questions should be addressed as, for example, the availability of massive amount of data for training complex algorithms, the reliability and dependability of the predictions, the required computational resources, and the acceptability of the final outcomes by the users.

Nevertheless, the joint effort of psychologists, engineers, and health care specialists can provide, in a near future, new affordable \gls{iot}-based \gls{aal} solutions that can fully meet all the requirements of elderly and fragile people.

\section*{Acknowledgment}
The authors acknowledge the economical support by the POR FESR 2014-2020 Work Program of the Veneto Region (Action 1.1.4) through the project No. 10066183 titled \emph{"Sistema domotico IoT integrato ad elevata sicurezza informatica per smart building"}.
Part of this work was also supported by MIUR (Italian Minister for Education) under the initiative \emph{"Departments of Excellence"} (Law 232/2016).

\bibliographystyle{IEEEtran}
\bibliography{ref}

\end{document}